\documentstyle[12pt,cite,epsf,epsfig]{article}
\newcommand{\newc}{\newcommand}
\newc{\tev}{\,{\rm TeV}}
\newc{\gev}{\,{\rm GeV}}
\newc{\sgn}{\mr{sgn}\,}
\newc{\ra}{\rightarrow}
\newc{\rpv}{$\mathrm{\not\!R_p}$}
\newc{\met}{$\not\!\!E_T$}
\newc{\rp}{$\mathrm{R_p}$}
\newc{\real}{\mathcal{R}e}
\newc{\alsm}{{\displaystyle \sum_{\alpha=1,2}}}
\newc{\besm}{{\displaystyle \sum_{\beta=1,2}}}
\newc{\al}{\alpha}
\newc{\ga}{\gamma}
\newc{\de}{\delta}
\newc{\cw}{\cos\theta_w}
\newc{\ssw}{\sin^2\theta_w}
\newc{\ccw}{\cos^2\theta_w}
\newc{\cbe}{\cos\beta}
\newc{\sbe}{\sin\beta}
\newc{\sh}{\hat{s}}
\newc{\sa}{\sin\al}
\newc{\ca}{\cos\al}
\newc{\bv}{$\mathrm{\not\!B}$}
\newc{\lv}{$\mathrm{\not\!L}$}
\newc{\ie}{{\it i.e.\/}\ }
\newc{\lam}{\lambda}
\newc{\cht}{\tilde{\chi}}
\newc{\upt}{\tilde{u}}
\newc{\elt}{\tilde{\ell}}
\newc{\hgt}{\tilde{H}}
\newc{\nut}{\tilde{\nu}}
\newc{\dnt}{\tilde{d}}
\newc{\psb}{\bar{\psi}}
\newc{\rtt}{\sqrt{2}}
\newc{\mut}{\tilde{\mu}}
\newc{\mr}{\mathrm}
\newc{\bath}{\bar{\theta}}
\newc{\tht}{\theta}
\newc{\JC}{{\bf J}}
\newc{\lra}{\longrightarrow}
\newc{\eg}{{\it e.g.\,}}
\newc{\barr}{\begin{array}}
\newc{\earr}{\end{array}}
\newc{\bary}{\begin{eqnarray}}
\newc{\eary}{\end{eqnarray}}
\newc{\dis}{\displaystyle}
\newc{\beq}{\begin{equation}}
\newc{\eeq}{\end{equation}}
\newc{\me}{\mathcal{M}}
\newc{\dbm}{\partial_\mu}
\newc{\sgm}{\sigma_\mu}

\def\ra{\rightarrow}
\def\dis{\displaystyle}
\def\bentarrow{\:\raisebox{1.1ex}{\rlap{$\vert$}}\!\rightarrow}

\def\dk#1#2#3{
        \begin{equation}
        \begin{array}{r c l}
        #1 & \rightarrow & #2 \\
         & & \bentarrow #3
        \end{array}
        \end{equation}
                }

\catcode`@=11 
\def \gsim{\mathrel{\mathpalette\@versim>}}
\def \lsim{\mathrel{\mathpalette\@versim<}}
\def \@versim#1#2{\lower0.4ex\vbox{\baselineskip\z@skip\lineskip\z@skip
     \lineskiplimit\z@\ialign{$\m@th#1\hfil##\hfil$%
     \crcr#2\crcr\sim\crcr}}}
\catcode`@=12 
\def\gev{\: \rm GeV}

\begin{document}
\setcounter{page}{0}
\renewcommand{\thefootnote}{\fnsymbol{footnote}}
\thispagestyle{empty}

\begin{titlepage}
\begin{flushright}
\end{flushright}
\vspace{+2cm}

\begin{center}
{ \Large \bf NLO QCD corrections to excited lepton production at the LHC}\\
\vspace{+1cm}

\bf{Swapan Majhi}\footnote{E-mails:  tpskm@iacs.res.in}\footnote{Work supported
by CSIR Pool Scheme (Pool No. 8545-A)}\\

\it{Department of Theoretical Physics, \\
Indian Association for the Cultivation of Science \\
Kolkata 700032 India.}
\end{center}
\setcounter{footnote}{0}
\begin{abstract}\noindent
We revisited excited leptons ($\bar{l^*}l$)  
production through gauge mediation only at LHC, 
followed by their two body decays 
into Standard Model (SM) particles.
We include the next-to-leading order (NLO)
QCD corrections to these processes. 
We have shown that these corrections can be substantial 
and significant. 
We also show that the scale dependence of the NLO cross section is 
greatly reduced as compare to leading order (LO) cross section.
\end{abstract}
\end{titlepage}

\setcounter{footnote}{0}
\renewcommand{\thefootnote}{\arabic{footnote}}

\section{Introduction}

%

     In the recent years in particle phenomenology, a lot of attention goes into
find the suitable model beyond the standard model (BSM) physics, which can 
explain many issues like the replication of the fermion families, dark matter, 
baryogenesis etc. that are still not understood within the framework of the 
Standard Model (SM).
The quark-lepton composite model \cite{llmodel} is one of the prime 
candidate among others 
like supersymmetry~\cite{susy}, grand unification~\cite{Pati:1974yy,GUTS} 
(with or without supersymmetry), family symmetries (gauged or otherwise).

The replication of fermion generations suggests the possibility of
quark-lepton compositeness. In these theories\cite{comp_mod, Additional_comp}, 
the fundamental constituents, {\em preon}s\cite{preon}, 
experience an additional strong and confining force. 
At energies far above a certain (compositeness) scale ($\Lambda$),
preons are almost free.
Below this scale $\Lambda$, 
the interaction of preons become very strong forcing them to form a bound state 
of quarks and leptons.
Understandably, in such models, higher (excited) states
          of quarks ($q^*$) and leptons ($l^* $) must also exist.

Since the composite fermion is just an excited state of the SM fermion,
generalized dipole moment-like terms should mediate interactions between
them. The correspondng effective Lagrangian\cite{LagrangianGM} is given by
\beq
{\cal L}_{GM} = {1 \over 2 \Lambda} \bar{f^*_R} \sigma^{\mu\nu}
\Big[g_s f_s {\lambda^a \over 2} G^a_{\mu\nu}
+g f'' {\tau \over 2}.W_{\mu\nu} + g' f'{Y\over 2} B_{\mu\nu}
\Big]f_L + h.c.
     \label{lagrangianGM} 
\eeq
where $G_{\mu\nu}^a, W_{\mu\nu}$ and $B_{\mu\nu}$ are the field strength tensor
of the $SU(3)$, the $SU(2)$ and the $U(1)$ gauge fields respectively. 
$f^*$ and $f$ denote the excited fermion and SM fermion respectively.
$f_s,f''$ and $f'$ are the parameters of the compositeness. 
Usually they are taken to be order of 1.

In the absence of a full theory, all the interactions of such composite
fermions cannot be written down unambiguously. Rather, one must
take recourse to an effective Lagrangian. The latter, typically, would contain
not only the term of eqn.(1) above but others as well. The corresponding
Wilson coefficients can only be determined if the ultraviolet completion
was well-known and are,  a priori, unknown within the context of the
effective theory. The literature abounds with the discussion of one such subset
of operators, namely four-fermion interactions between a pair of SM fermions
and a pair of composites (the so-called contact 
interaction~\cite{LagrangianCI }). It should
be realized that such operators are suppressed by an additional power of
$\Lambda$ as compared to the terms of eqn.(1). Thus, it makes eminent sense
to consider the above while neglecting the four-fermion terms.

It is evident that these operators may lead to 
significant phenomenological effects in collider experiments, 
like $e^+ e^-$~\cite{Delphi}, $e \, P$~\cite{HERA2} or 
hadronic\cite{cdfprl,CMS_CI,ATLAS_CI}. 
It is quite obvious that
the effects would be more pronounced at higher energies,  given the
higher-dimensional nature of ${\cal L}_{GM}$. 
The best low-energy bounds on such composite operator would arise
from the precise measurement of leptonic branching ratios (BR)
of lepton $\tau$\cite{TauBR}. The loop effect of these excited
states can modify the SM branching ratio predictions and comparison 
with the experimental data can impose bounds on masses of these new particles
and their couplings. These bounds are quite weak \cite{TauBR_bounds}.
The constraints on such excited states 
came from the Delphi~\cite{Delphi} and 
CDF~\cite{cdfprl} experiments.
More recently, the 
measurement of the $\bar{l} l \gamma$ cross section\cite{CMS_CI, ATLAS_CI} 
at high invariant masses 
sets the most stringent limits on contact interactions. 

It is a well known fact that the QCD corrections can alter the cross sections
 quite significantly  
at hadron colliders.
Recently, the production of $\bar{l^*} l (\bar{l} l\gamma)$ in the context of
 contact interactions have received much attention
from both CMS\cite{CMS_CI} and ATLAS\cite{ATLAS_CI} collaborations.
They have searched for heavy excited lepton via $\bar{l} l\gamma$ channel 
and put the mass bound on excited lepton at center of mass energy 
($\sqrt{S} = 7$ TeV). 
Due to small production cross section of $\bar{l^*} l$ (and hence $\bar{l} l\gamma$)
 through gauge mediation 
(eqn.(\ref{lagrangianGM})) at $\sqrt{S} = 7$ TeV, 
they did not open up
this production channel. 
 They have analyzed their data based on the leading order calculation due to
non-existent of higher order calculations for this process. Recently, we have 
calculated NLO QCD corrections to this process\cite{swapan_excited} and CMS 
collaborations are analyzing their data again using our NLO result at $\sqrt{S} = 8$ TeV 
(in private communications). At higher center of mass energy ($\sqrt{S} = 13~ (33)$
 TeV) and/or high luminosity (HL-LHC), the production cross section of $\bar{l^*} l$ through
 gauge mediation are considerably large and may lead to important phenomenological 
consequences. 
This is our prime interest in this article.
However, there exists no higher order calculations for this process. 
While it may seem that the NLO QCD corrections to the processes driven by such
non-renormalizable interactions are ill-defined, 
it is not quite true\cite{SM_CI,ravi_gravi1}. In particular, 
if the interaction can be factorized into two currents such that one current
with colored object and other current with colored neutral object then
the NLO QCD corrections can be done with colored current one without any
difficulties. For example, Ref.\cite{SM_CI} dealt with contact interaction
with SM fermions. In this article, we 
have computed the NLO QCD corrections to the processes mentioned above.

The rest of the article is organized as follows. 
In Section \ref{NLO corrections}, we 
start by outlining the general methodology and follow it up with the 
explicit calculation of the NLO corrections to the differential 
distribution in the dilepton ($\bar{l^*}l$)  invariant mass.  Section \ref{results} we present our numerical results. And finally, we summarize in Section \ref{conclusion}.

\section{NLO corrections}
\label{NLO corrections}
        We reconsider excited leptons production 
through gauge mediated interaction as exemplified by 
eqn(\ref{lagrangianGM}) at LHC. 
The processes are

\dk{P(p_1)+P(p_2)}{{l^{*}}^{\!\!\!\!\!\!^{^{(-)}}}(l_1)+ {\bar{l}}^{\!\!\!^{^{^{^{(\,\, )}}}}}(l_2) + X(p_X)}{l^{\!\!\!\!\!\!^{^{(-)}}}(l_3)+V(p_4)\;,}
where $p_i(i=1,2)$ denote the momenta of the incoming hadrons and 
$l_i$ are those for the outgoing leptons. 
Similarly, the momentum $p_X$ carries by the inclusive hadronic state $X$.
The $p_4$ is the outgoing 
vector boson's $V(V = \gamma,Z,W^{\pm})$ momentum.
We have considered only two body leptonic decay of excited leptons for 
our interest.
The hadronic cross section is defined in terms of the partonic cross section
 convoluted with the appropriate parton distribution functions $f_a^P(x)$
and is given by 
\beq
2 S {d\sigma^{P_1 P_2} \over d Q^2 } = \sum_{ab= q,\bar{q},g} \int_0^1 dx_1\: 
\int_0^1 dx_2 \: f^{P_1}_a(x_1) \: f^{P_2}_b(x_2) \, 
\int_0^1 dz ~2\, \hat{s}\; {d\sigma^{a b } \over d Q^2} \, 
\delta(\tau - z x_1 x_2),
\label{eq:hadr_cross1}
\eeq
where $x_i$  is the fraction of the initial state proton's momentum 
carried by the $i^{th}$ parton. i.e. the parton momenta
$k_i$ are given by $k_i = x_i \,  p_i$.
The other variables are defined
as
\beq
\barr{rclcrclcrcl}
S &\equiv& (p_1+p_2)^2
& \qquad &
\hat{s} &\equiv& (k_1+k_2)^2
 & \qquad &
Q^2 &\equiv&  (l_1+l_2)^2
\\[2ex]
\tau&\equiv& \dis {Q^2 \over S}
& \qquad &
 z &\equiv& \dis {Q^2 \over \hat{s} }
& \qquad &
\tau &\equiv& z\,x_1\,x_2 \ .
\earr
\eeq
Although the
effective Lagrangian is a non-renormalizabe one,
due to its current-curent structure
the offending higher order QCD corrections 
possible\cite{SM_CI,ravi_gravi,ravi_gravi1}.
Since the QCD corrections affect only hadronic currents with leptonic 
current being a mute spectator, the offending higher dimensional nature of
of the effective Lagrangian never comes into play in our calculations.
Therfore it is convenient to express our matrix element 
for the process as a sum of several current-current pieces with 
a ``propagator'' in between.
In other words, symbolically,
\beq
{\cal M}^{\rm Total} = \sum_j \, {\cal J}^{\rm Had}_j
                        \cdot  P_j \cdot {\cal J}^{\rm Lept}_j
\eeq
where the dots ($\cdot$) denote Lorentz index contractions as appropriate
and the propagators $P_j$ are
\beq
\barr{rclcl c rclcl}
P_{\gamma} &=& \dis {i \over Q^2} g_{\mu \nu}
           & \equiv & g_{\mu \nu} \tilde{P}_{\gamma}
& \qquad &
P_Z &=& \dis {i \, g_{\mu \nu} \over Q^2 - M_Z^2 - i M_Z \, \Gamma_Z }
& \equiv &  g_{\mu \nu} \tilde{P}_Z .
\earr
\eeq
With this definition,
the hadronic cross section can be written as
\beq
2 S {d\sigma^{P_1 P_2} \over dQ^2}(\tau, Q^2) = {1 \over 2 \pi}
\sum_{j,j^{\prime}= \gamma, Z}
\tilde{P}_j(Q^2) \: \tilde{P}_{j^{\prime}}^{*}(Q^2) \:{\cal L}_{jj^{\prime}}(Q^2)
\: W_{jj^{\prime}}^{P_1 P_2}(\tau, Q^2)
\eeq
where the hadronic structure function $W$ is defined to be
\beq
W_{jj^{\prime}}^{P_1 P_2}(\tau, Q^2) = \sum_{a,b,j,j^{\prime}} \int_0^1 dx_1\:
\int_0^1 dx_2 \: f^{P_1}_a(x_1) \: f^{P_2}_b(x_2)
\int_0^1 dz ~ \delta(\tau - z x_1 x_2)
\bar{\Delta}^{jj^{\prime}}_{ab}(z, Q^2, \epsilon) \ .
\eeq
Note that, the bare partonic coefficient function
$\bar \Delta$ contains all the  singularities, namely,
ultraviolet, soft and collinear divergences. To handle these, 
we have followed the
dimensional regularization ($DR$) scheme. The renormalization of
$VA$-type interactions is quite established (see for example,
in Ref.\cite{ravi_gravi}). After the renormalization,
one must get the ultraviolet regularized (and renormalized)
expressions.
To the ultraviolet regularized expressions, we
must add the contribution from the real gluon emission processes 
(bremsstrahlung) as well as the Compton processes (gluon initiated processes).
In this way, we remove the soft singularities and the left over expressions
contain only collinear singularities.
These collinear singularities can be removed through mass factorization.
Finally one gets the finite coefficient function $\Delta$ as in 
eqn(\ref{dsig:dm}).

The leptonic tensor is given by
\beq
{\cal L}^{jj^{\prime} \rightarrow \,l\, l^{\prime} } =
\int \prod_i^{2}\Bigg({d^nl_i \over (2\pi)^n} \: 2 \pi
\,\delta^{+}(l_i^2)\Bigg)
(2\pi)^n\, \delta^{(n)}\Big(q - l_1 - l_2\Big)
|{\cal M}^{jj^{\prime}\rightarrow \,l^+ l^-}|^2 \ ,
\eeq
which leads to
 \beq
 {\cal L}_{jj^{\prime} \rightarrow l\, l^{\prime} }
    = \barr{rcl}
       \dis\Big( -g_{\mu \nu} + {q_{\mu} q_{\nu} \over Q^2} \Big) \,
                 {\cal L}_{jj^{\prime}}(Q^2)  & \qquad &
                    (j, j^{\prime} = \gamma,  Z) 
        \earr
\eeq
 with
 \beq
\barr{rcl c rcl}
 {\cal L}_{\gamma \gamma}(Q^2) &=& \dis {\alpha \over 6 \Lambda^2} \,
|f_{\gamma }|^2 \, Q^2 \,{\cal L}(Q^2)
& \qquad &
 {\cal L}_{Z Z}(Q^2) & = & \dis {\alpha \over 6 \Lambda^2} \,
|f_{Z}|^2 \, Q^2 \,{\cal L}(Q^2)
\\[2ex]
 {\cal L}_{\gamma Z}(Q^2) &=& \dis{\alpha \over 3 \Lambda^2 } \, 
 f_{Z}\,f_{\gamma } \, Q^2 \,{\cal L}(Q^2)
& &  {\cal L}(Q^2) & =  & Q^2 +m_1^2+m^2_2-{2 \over Q^2} \big(m_1^2-m_2^2\big)^2
\earr
\eeq

The physical hadronic cross section can be obtained
by convoluting the finite coefficient functions with appropriate 
parton distribution functions and hence the inclusive differential 
cross section is given by 
\beq
\barr{rcl}
\dis 2 S {d\sigma^{P_1 P_2} \over dQ^2 }(\tau, Q^2) &=& \dis
\sum_q \int_0^1 dx_1\: 
\int_0^1 dx_2 
\int_0^1 dz ~ \delta(\tau - z x_1 x_2) 
\dis {\cal F}^{VA}_q \; {\cal G}_{VA} 
\\[3ex]
{\cal G}_{VA} & \equiv & 
H_{q\bar{q}}(x_1,x_2,\mu_F^2)\Big\{\Delta^{(0),VA}_{q \bar{q}}(z,Q^2,\mu_F^2)
+ a_s \Delta^{(1),VA}_{q \bar{q}}(z,Q^2,\mu_F^2) \Big\} 
\\[2ex]
&+ &
\Big\{H_{q g}(x_1,x_2,\mu_F^2) + H_{gq}(x_1,x_2,\mu_F^2)\Big\} a_s
\Delta^{(1),VA}_{q g}(z,\mu_F^2),
\label{dsig:dm}
\earr
\eeq
where the renormalized parton flux $H_{ab}(x_1,x_2,\mu_F^2)$ and 
the finite coefficient functions $\Delta^{(i)}_{ab}$ are given 
in Refs.\cite{ravi_gravi,ravi_gravi1,SM_CI}.
The constant ${\cal F}^{VA}$ contains information of all the couplings, propagators and the massive final state particles which is given by 

\bary
{\cal F}^{VA}_q &=& {2\alpha^2\over 3}{\beta \over \Lambda^2} {\cal L}(Q^2)\Bigg[
|f_{\gamma}|^2 {e_q^2\over Q^2} - e_q f_{\gamma} f_{Z}\Big(g_q^L+g_q^R\Big) {(Q^2-m_Z^2)\over Q^2} Z_Q
\nonumber \\
&& + {1\over 2}|f_{Z}|^2 \Big((g_q^L)^2+(g_q^R)^2\Big) Z_Q
\Bigg],
\eary
\beq
\barr{rcl c rcl}
Z_Q &=& {Q^2 \over (Q^2-m_Z^2)^2 + \Gamma_Z^2 m_Z^2} 
& \qquad &
g_q^R &=& -2 T^3_q \csc\theta_W-e_q\tan\theta_W
\\[2ex]
g_q^L &=&-e_q\tan\theta_W \hspace{0.5cm}
& & \beta &=& \Bigg(1 + {m_1^4\over Q^4}+ {m_2^4\over Q^4} - 2{m_1^2\over Q^2}
- 2{m_2^2\over Q^2} - 2{m_1^2\over Q^2}{m_2^2\over Q^2}\Bigg)^{1\over 2}.
\earr
\eeq
\section{Results and Discussion}
\label{results}

In the previous section, we have calculated the differential distributions 
with respect to invariant mass ($Q$) of $\bar{l^*}l$ (one excited lepton and SM lepton).
For our interest, we have expressed the above differential 
distribution (eqn.(\ref{dsig:dm})) to the total cross section by integrating over $Q^2$ 
and it is given by
\beq
\dis \sigma^{P_1 P_2} (M^2_{*},S,\Lambda) = 
\int {d\sigma^{P_1 P_2} (\tau,Q^2)\over d Q^2} d Q^2 .
\label{eq:Lsl_cs}
\eeq

 In our numerical analysis, we present our results at three different LHC
energies $\sqrt{S}= 13, 33, 100$ TeV for the simplest case where
the factorization ($\mu_F$) and renormalization scale ($\mu_R$) considered 
to be equal to the invariant mass ($Q$) of $\bar{l^*}l$. 
Later, we have shown the scale dependence of our results by introducing
the factorization scale, $\mu_F^2 \,( \mu_R^2)\neq Q^2$.
Since the QCD correction
does not depend on the contact interaction scale $\Lambda$, for definiteness
we have used a particular value of $\Lambda = 2, 6$ TeV for each LHC energy 
unless it is quoted.
Through out our numerical analysis,
we have used Cteq6Pdf\cite{CTEQ6} and MSTW 2008 \cite{MSTW} parton distribution 
functions (PDFs) 
otherwise mentioned specifically. 
\begin{figure}[htb]
\centerline{
\epsfxsize=18cm\epsfysize=11cm
                     \epsfbox{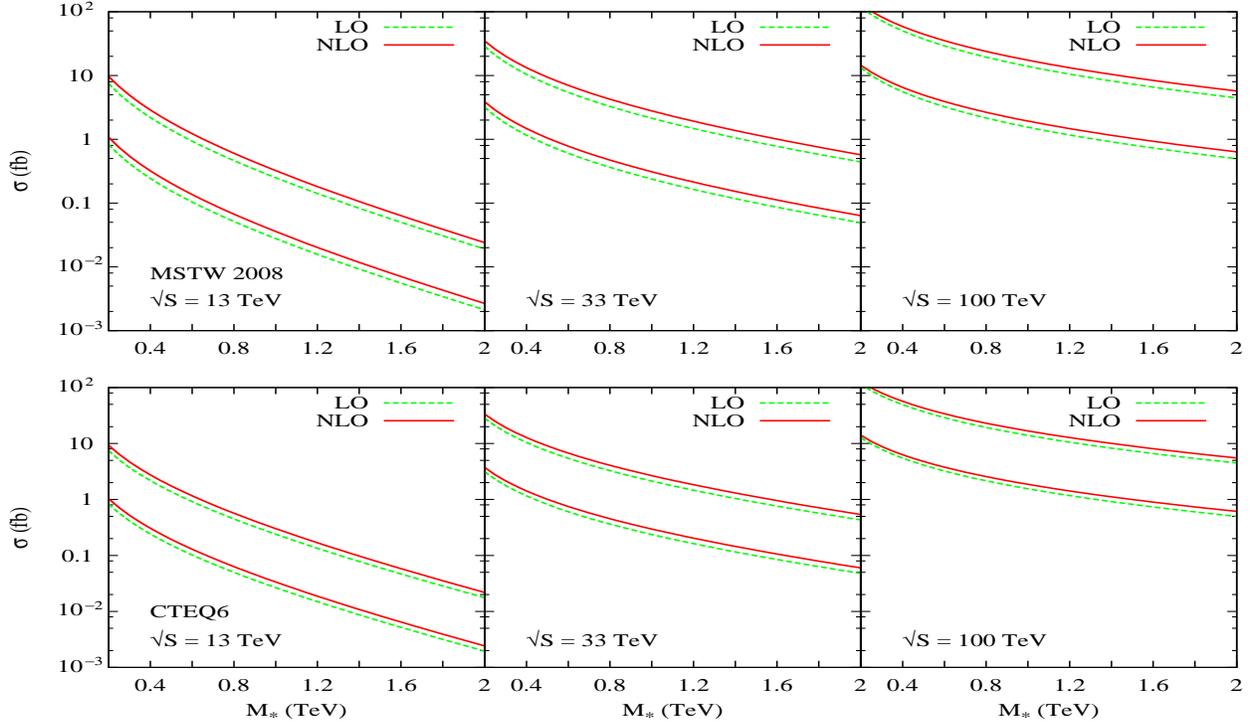}
}
\vspace{-1cm}
\caption{\em Variation of total cross-section for $\bar{l^*} l$
production with respect to
excited lepton mass ($M_*$) at the LHC.
        Upper (lower) set represents for $\Lambda = 2 (6)$ TeV.
        }
\label{fig:Ls_cs}
\end{figure}

We will first discuss the NLO corrections of 
$\bar{l^*} l$ (and $\bar{l} l^*$ as well)
productions in general and 
later we consider only a particular process $\bar{l} l \gamma$ production.
This particular process has been analyzed by both CMS\cite{CMS_CI} and 
ATLAS\cite{ATLAS_CI} in the context of contact interaction at 
low center of mass energy ($\sqrt{S} = 7$ TeV). 
Since the production cross section of $\bar{l^*} l$
through gauge mediation at this center of mass energy ($\sqrt{S} = 7$ TeV) 
is quite small, they did not open up this channel.
However at higher LHC energies
($\sqrt{S} = 13, 33$ TeV and/or 100 TeV), this particular process may 
play an important role for searching the excited lepton in the beyond standard
model scenario.
The above mentioned particular process attains through two body decaying process
of excited lepton ($l^*$).
The two body decay of 
excited lepton does not have any effect on QCD correction thus 
the NLO QCD correction to the $\bar{l} l \gamma$ production process is same
as NLO QCD correction to the $\bar{l^*} l$ as described in the next 
section \ref{llg}.

\begin{figure}[htb]
\centerline{
\epsfxsize=14cm\epsfysize=7cm
                     \epsfbox{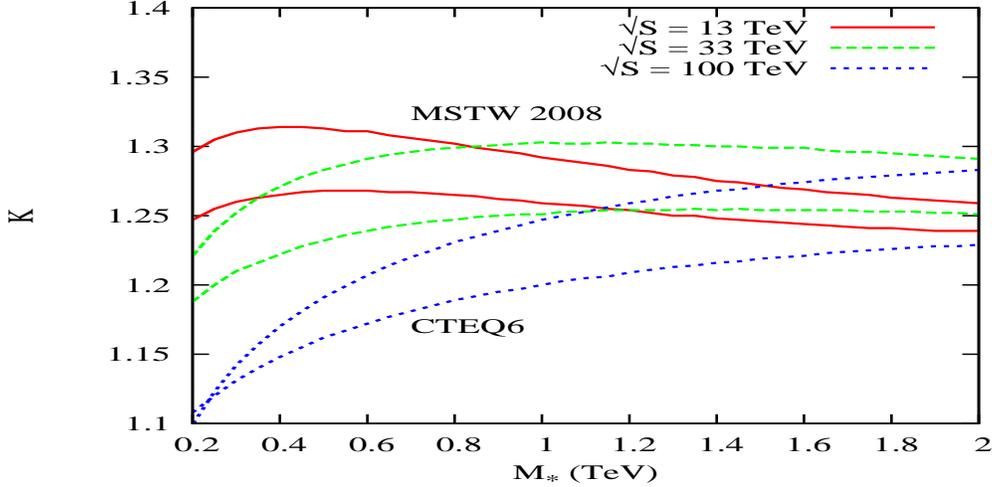}
}
\vspace{-0.5cm}
\caption{\em Variation of $K$-factor with respect to the excited lepton
mass ($M_*$) for $\Lambda = 2$ TeV at the LHC for $\bar{l^*}l$ channel.
Upper (lower) set is for MSTW 2008 (CTEQ6) PDFs.
        }
\label{fig:KLs}
\end{figure}

In figure \ref{fig:Ls_cs}, we have plotted the total cross section of ($\bar{l^*}l$)
as a function of excited lepton mass ($M_* $). The cross sections presented
in figure \ref{fig:Ls_cs} contain the contributions of all the light flavors 
($u,d,s$-quarks) as those for heavier flavor being essentially negligibly small.
As we have seen from figure \ref{fig:Ls_cs} that 
the cross section decreases with  
excited lepton mass ($M_* $) due to not only the fall of partonic cross section but also
due to the fall in parton distribution functions (and hence effective flux of $q\bar{q}$ as well as $qg$)
at higher momentum fraction ($\tau$ and hence Bjorken scale $x$). 
The fall of the total cross section is more at lower center of mass (c.o.m.) 
energies 
than the higher c.o.m. energy. 
At higher momentum fraction $\tau$, we are integrating 
over small phase space 
region at low center of mass energy $\sqrt{S}$.

   To quantify the enhancement of NLO cross section, we define a variable
called $K$-factor as given by
\beq
K = {\sigma^{NLO} \over \sigma^{LO}}
\eeq
where the LO (NLO) cross sections are computed by convoluting the corresponding 
parton-level cross sections with the LO (NLO) parton distribution functions.

In figure (\ref{fig:KLs}) we have 
shown the variation of $K$-factor with respect to the 
excited lepton mass ($M_*$). 
The variation of the total $K$-factor is about $25\%-30\%$  
 for c.o.m energies $\sqrt{S} = 13,\, 33$ TeV. 
For much higher c.o.m energy $\sqrt{S} = 100$ TeV,
the variation of the total $K$-factor is about $10\% - 25\%$
for $M_* \leq 1$ TeV and increases very fast.
For for $M_* \geq 1$ TeV, the $K$-factor increases slowly with $M_*$
($20\% - 30\%$).
In figures (\ref{fig:KLs}), the rate of fall of the
$K$-factor is much slower at lower
c.o.m energy (say $\sqrt{S} = 13,\, 33$ TeV) than the higher c.o.m energy $\sqrt{S} = 100$ TeV. 
At lower c.o.m energy (higher momentum fraction, $0.00024 \leq \tau \leq 0.024$ for $\sqrt{S} = 13$ 
and hence the Bjorken $x$), 
we are integrating relatively smaller phase space region.
As excited lepton mass increases, the variation of $K$-factor becomes smooth due to
the fact that the valence quark (mostly $u$ and $d$-quark) distributions dominate over 
sea quark and gluon distributions. 
At higher c.o.m energy (lower momentum fraction, $0.000004 \leq \tau \leq 0.00004$ for
$\sqrt{S} = 100$ TeV),                         
we are integrating over relatively larger phase space. In this region,
the sea ($s$) quark and gluon ($g$) distributions dominate over 
the valence quark ($u$ and $d$-quark) distributions.
Therefore the compton-like subprocess 
(in particular, $g\, s(\bar{s})\rightarrow \bar{l^*}\, l\, s(\bar{s})$) dominates due to
large gluon ($g$) and sea-quark fluxes and hence explains such behavior 
of $K$-factor. 
 
\subsection{$\bar{l}l\gamma$ production}
\label{llg}
The decaying of heavy excited lepton into a light SM lepton and a electroweak
gauge bosons $V(\equiv \gamma, Z, W)$ according
to the Lagrangian(\ref{lagrangianGM}) produce a particular process 
($\bar{l}l\gamma$) of our prime interest.
Therefore the total NLO cross section of lepton pair ($\bar{l}l $) 
and a gauge boson $V$ 
can be calculated by multiplying the branching ratio to the 
eqn.(\ref{eq:Lsl_cs})  as given below 
\beq
\dis \sigma^{P_1 P_2} (M^2_{*},S,\Lambda) = BR(l^*\rightarrow l V) 
\int {d\sigma^{P_1 P_2} (\tau,Q^2)\over d Q^2} d Q^2 .
\eeq
\begin{figure}[htb]
\vspace{-1cm}
\centerline{
\epsfxsize=18cm\epsfysize=14cm
                     \epsfbox{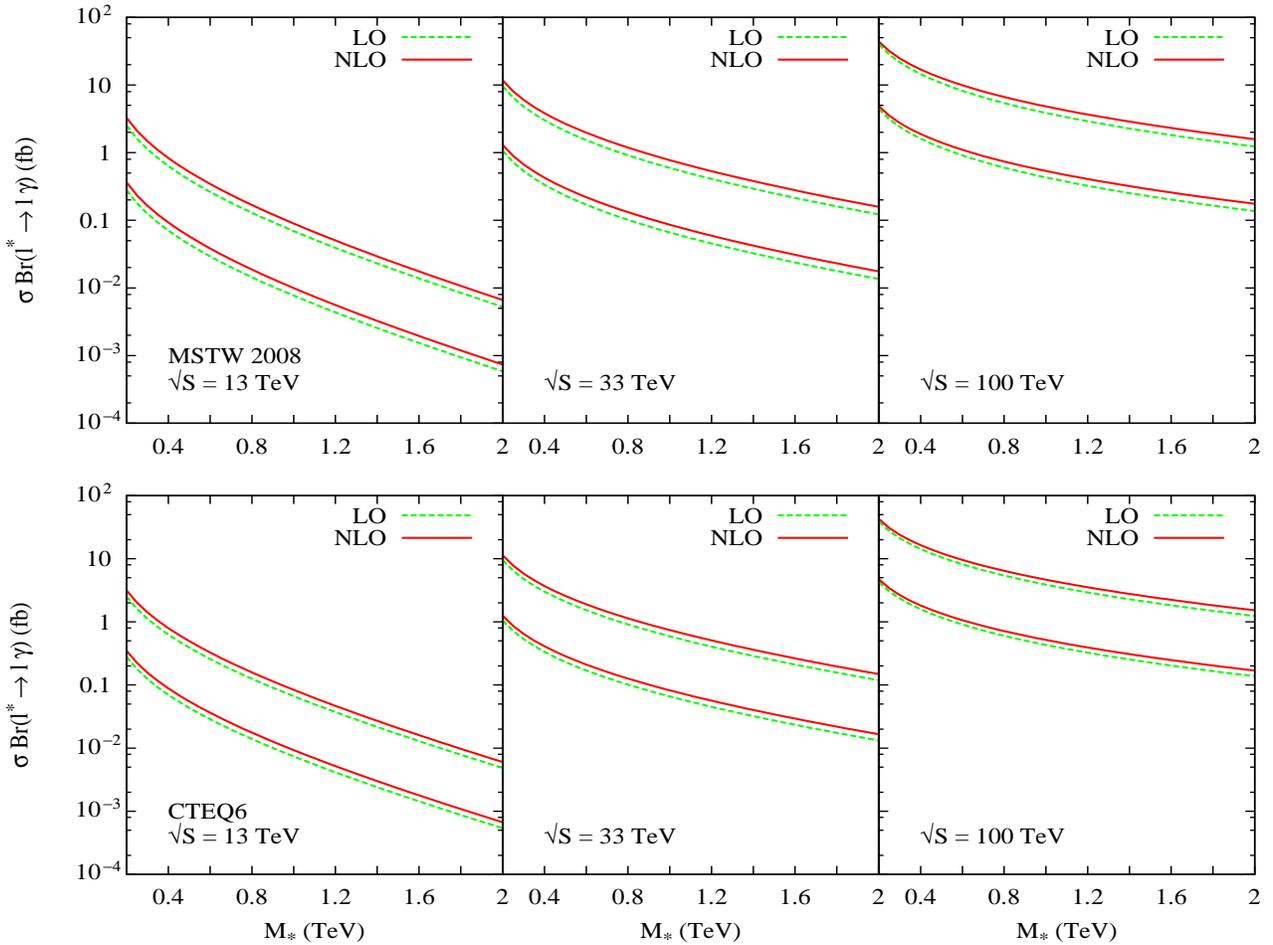}
}
\vspace{-1.0cm}
\caption{\em Total cross-section for $l \bar{l}\gamma$ production
             at the LHC.
        For each set, the solid (dashed) lines refer to NLO (LO) cross
        sections. Upper (lower) set is for $\Lambda = 2 (6)$ TeV.
        }
\label{fig:llg_cs}
\end{figure}

The partial decay width of excited lepton for various 
electroweak gauge bosons is given by
\beq
\Gamma(l^*\rightarrow l V) = {1\over 8} \alpha \,f^2_V {M_*^2 \over \Lambda^2} \Bigg(1-{m^2_V\over M^2_*}\Bigg) \Bigg(2+{m^2_V\over M^2_*}\Bigg),
\eeq
with
\bary
f_{\gamma} &=& f\,T_3+f'{Y\over 2},\\
f_{Z} &=& f\,T_3\cot\theta_W-f'{Y\over 2}\tan\theta_W,\\
f_{W} &=& {f\over \sqrt{2}}\csc\theta_W, 
\eary
where $T_3$ denotes the third component of the weak isospin and $Y$ represents
the weak hypercharge of excited lepton. $\theta_W$ is the Weinberg's angle.
The compositeness parameters $f$ and $f'$ are taken to be unity through
out our analysis. The variation of these parameters have been considered 
in elsewhere (for example in the Refs.\cite{PRD65,PRD81}). The details of decay width
and branching fraction of excited lepton is given in 
\cite{swapan_excited} (see table 1 and references therein).

\begin{figure}[htb]
\centerline{
\epsfxsize=14cm\epsfysize=8.0cm
                     \epsfbox{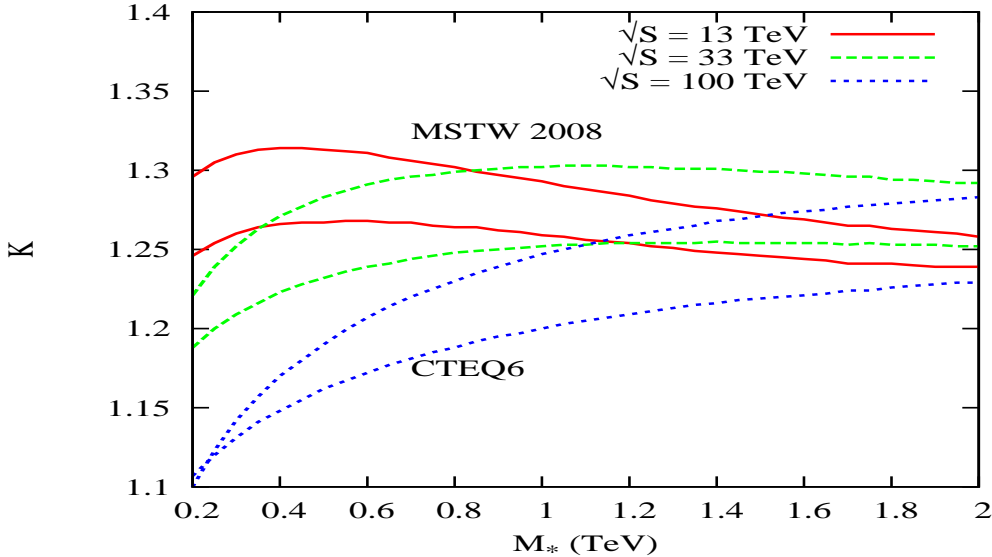}
}
\vspace{-0.5cm}
\caption{\em $K$-factor for $l \bar{l}\gamma$ production
             at three different LHC energies.
Upper (lower) set is for MSTW 2008 (CTEQ6) PDFs.
        }
\label{fig:ktot}
\end{figure}

In figure \ref{fig:llg_cs}, we have plotted the total cross section versus 
invariant mass $M_* (\equiv M_{l\gamma})$ of one SM lepton($l$) and a 
photon ($\gamma$). 
We have shown for two different PDFs, namely, CTEQ6\cite{CTEQ6} and 
MSTW 2008\cite{MSTW}. 
As explain before, the cross section decreases in increase of 
invariant mass $M_*$.
The variation of the cross section looks same 
for two different PDFs, actually they are not. This can be
found out from figure \ref{fig:ktot} and has been explained later on.
From the figure \ref{fig:llg_cs},                              
we see that as the composite scale ($\Lambda$) increases, the cross
section (both LO as well as NLO) decreases uniformly as $\Lambda^{-2}$ as 
expected (from eqn.(\ref{lagrangianGM})) for a
fixed center of mass energy ($\sqrt{S}$) and for different values of $\Lambda$,
the cross section scales accordingly. Therefore, one can obtain the
 cross section (for both LO as well as NLO) for arbitrary values of 
$\Lambda$ by multiplying with an appropriate scale factor to our results.

\begin{figure}[htbp]
\centerline{
\epsfxsize=18cm\epsfysize=20cm
                     \epsfbox{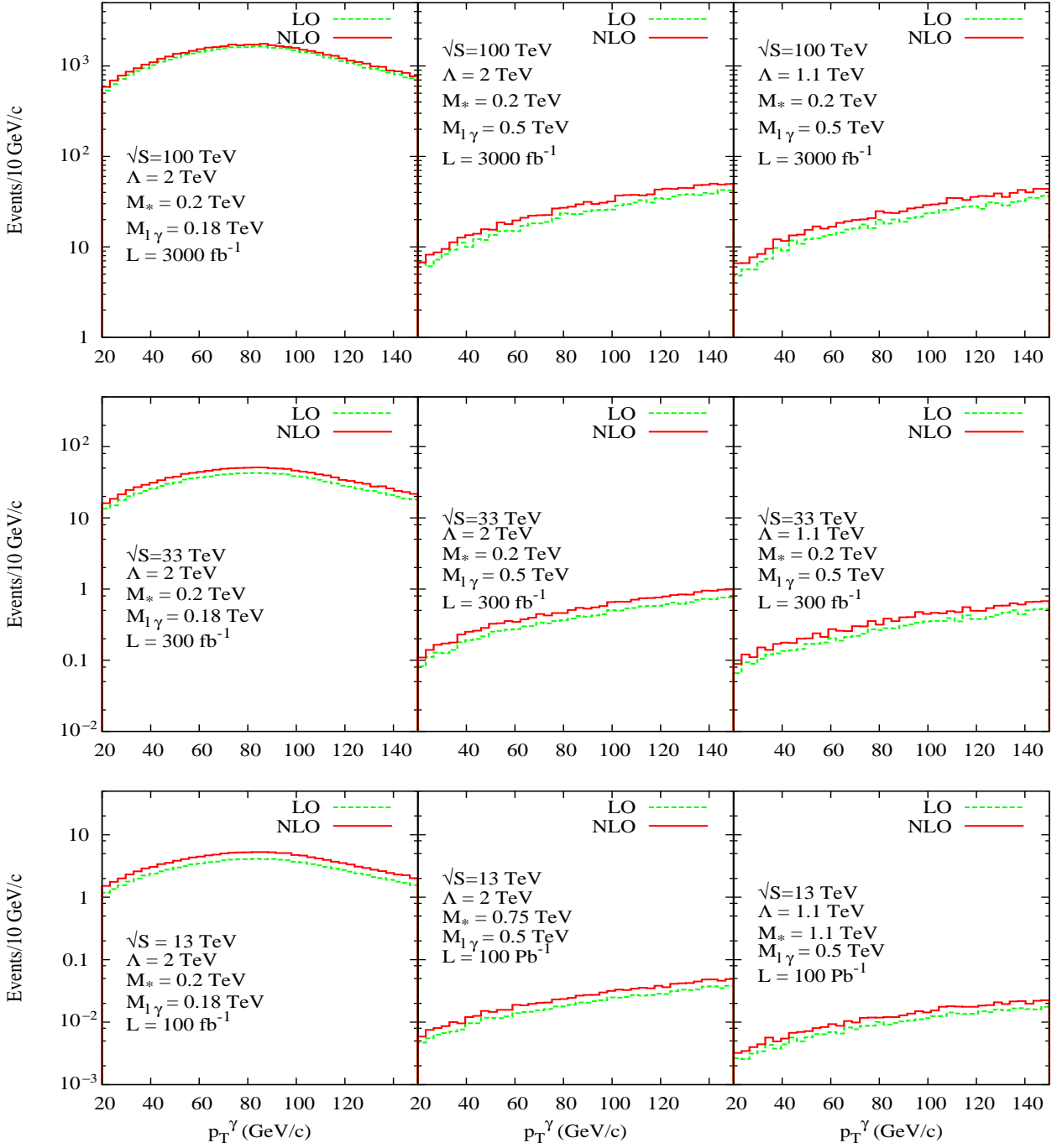}
}
\vspace{-1cm}
\caption{\em Photon transverse momentum distributions at three different
excited lepton masses and three different LHC energies for MSTW 2008 parton distribution functions.
        }
\label{fig:pT_dist2}
\end{figure}

In figures \ref{fig:pT_dist2}, we have displayed
the photon transverse momentum distribution. 
In this case, we use 
with same lepton-photon invariant mass cut ($M_{l\gamma}^{cut}$) as given
 in \cite{CMS_CI}. 
We have considered the projected luminosity 
100, 300 (3000) $fb^{-1}$ at $\sqrt{S} = 13, 33\, (100)$ TeV LHC energies respectively.
From the figures, one can see that the 
production rates are increased by including NLO QCD corrections.  

We have shown the variation of total $K$-factor with respect to the $M_*$
in figure \ref{fig:ktot}.
The variation of $K$-factor is very similar to the figure \ref{fig:KLs} as expected
and hence it has been explained there itself.
 The very wide range of $K$-factor difference between the two PDFs, namely, CTEQ6
and MSTW 2008 is due
to their different parameterizations of  their parton
distribution functions (owing to their use of different data sets
to extract the PDFs).

\subsection{The choice of Scale}

\begin{figure}[htb]
\vspace{-0.5cm}
\centerline{
\epsfxsize=18cm\epsfysize=14cm
                     \epsfbox{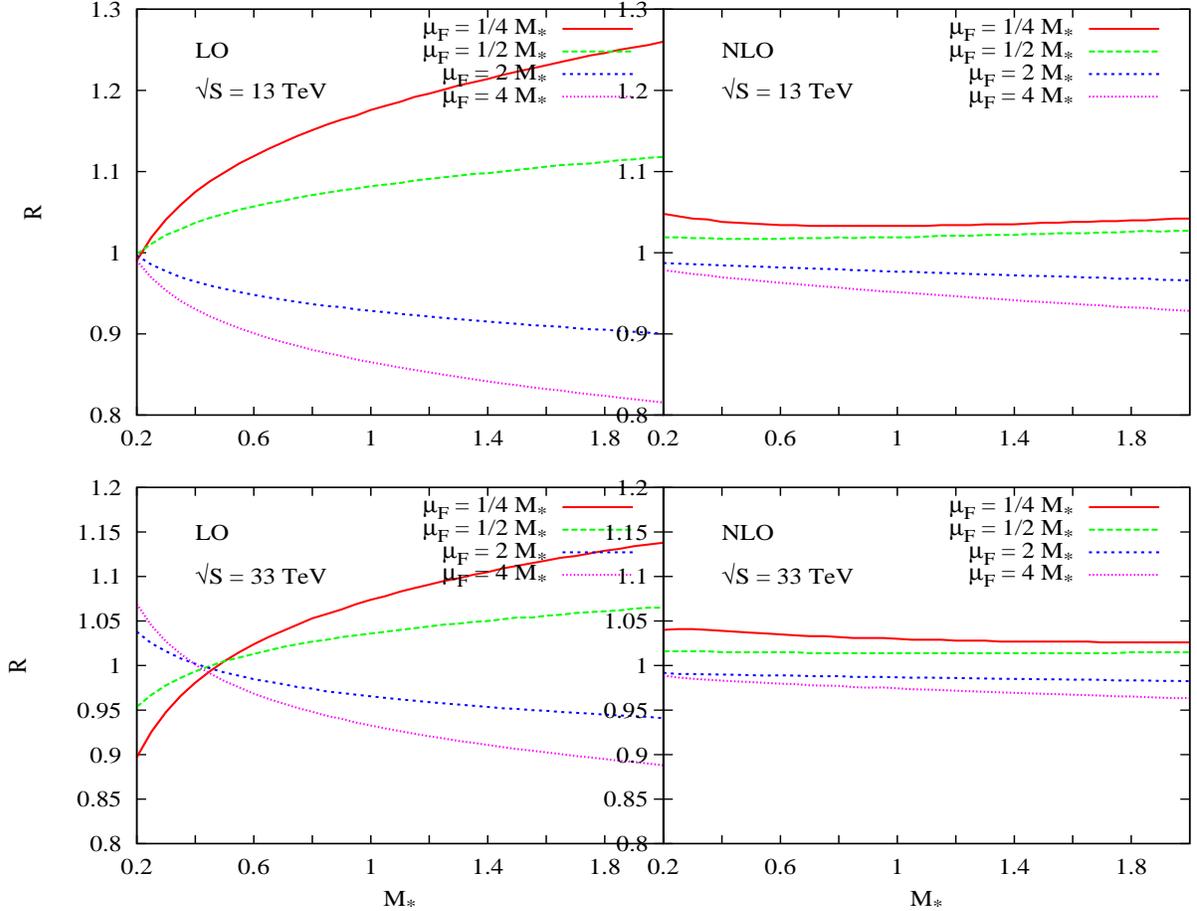}
}
\vspace{-1.5cm}
\caption{\em Variation of the ratio, R (defined in eqn.(\ref{ratio})) with respect to the excited lepton mass ($M_*$) at different factorization scale $\mu_F$ using CTEQ6 PDFs. 
        }
\label{fig:scal_var}
\end{figure}

         In our above discussions, we have considered 
the factorization scale, $\mu_F$, (relevant to both LO and NLO) and
the renormalization scale, $\mu_R$, (relevant only to NLO) to be same as 
the invariant mass $(Q)$ of $\bar{l^*} l$.
However the
cross section depends only on physical scales like 
the c.o.m. energy ($\sqrt{S}$)
and the masses of final state particles ($M_*$).
Since there is no theoretical guideline to choose a particular scale choice
the abovementioned scale choice is completely arbitrary.
Now we can check the scale dependence of our result by introducing 
another scale called factorization scale $\mu_F^2 ( = \mu_R^2$
the renormalization scale, for simplicity).
To quantify the scale dependence if we define
a ratio R,
\beq
R^I = {\sigma^I(S,M_*,\mu_F^2) \over \sigma^I(S,\mu_F^2=M^2_*)},~~~~ I = LO,~NLO,
\label{ratio}
\eeq
the ratio $R$ close to unity signify low sensitivity to scale choice and
hence a more robust result.
 
In figure \ref{fig:scal_var}, we have shown the variation of the cross scetion
with respect to the excited lepton mass at different the factorization scale.
From the figure \ref{fig:scal_var}, 
it is clear that the scale dependence reduces greatly at NLO cross section 
compare
to LO cross section. This signifies the necessity of NLO QCD corrections. The 
remaining very small scale ambiguity can be reduced by adding still higher
order corrections.

\section{Conclusions}
\label{conclusion}

In conclusion, we have systematically performed 
the next-to-leading order QCD corrections for the chromomagnetic type
interactions as given in eqn.({\ref{lagrangianGM}}). 
As opposed to naive expectations, we have showed that 
the QCD corrections are meaningful and reliable 
to such non-renormalizable theory.

We have analyzed the variation of cross section with respect to
the excited lepton mass (and hence the invariant mass of one SM lepton
and a SM gauge boson) at the LHC.
The enhancement of NLO cross section over the LO cross section is 
found to be quite significant. To quantify the enhancement, we present 
the corresponding $K$-factors  
in a suitable form 
for experimental analysis. 
We have also showed the scale dependence of our results. 
As expected, we have seen that
the scale dependences reduce greatly for the NLO 
results as compared to that for the LO case. 

\section*{Acknowledgments} 
Author would like to thank Debajyoti Choudhury for useful discussions and comments. 


\end{document}